# Using Gemini and LuaLaTeX to transcribe physics videos into PDF/UA-2 and ISO 32005 math-accessible PDFs

Craig W. Looney*
Christopher L. Duston

Department of Mathematics and Physics, Merrimack College, North Andover, MA, USA

August 20, 2026

**Abstract**

We present an AI transcription workflow that uses freely available tools to transcribe instructional physics videos with substantial mathematical content into math-accessible PDFs. Guided by our carefully crafted system instructions, Google Gemini transcribes linked or uploaded videos into LaTeX output. Importantly, all board-written or screen-presented equations are transcribed into LaTeX math and incorporated into the verbal flow or presented in descriptive segments, which also describe graphs, diagrams, and pedagogically important actions and interactions. Gemini's LaTeX output is appended to a prewritten SE tagging-enabled preamble and compiled by a LuaLaTeX engine; the resulting PDF transcripts routinely compile without errors and pass PDF/UA-2 and ISO 32005 validation tests. We provide a balanced performance assessment using numerous examples from test transcriptions. The publicly available supplementary resources contain complete transcripts corresponding to all presented examples, and also contain our well-tested system instructions and preamble so that others can use, adapt, study, and further develop our transcription framework.

## 1 Introduction

While well-made instructional videos have enhanced the general accessibility of mathematically intensive STEM content by expanding access to high-quality materials and by supporting effective interactive pedagogies, they present an accessibility challenge for blind and low-vision students. Legally and ethically, educational institutions face sweeping mandates to provide accessible materials under the Americans with Disabilities Act (ADA) [1] and international standards, such as the Web Content Accessibility Guidelines (WCAG) [2] and the unified PDF/UA-2 and ISO/TS 32005 accessibility standards [3]. However, translating a multi-channel, multi-sensory educational experience – where an instructor simultaneously speaks, gestures, and writes complex mathematics – into a single-channel, screen-readable text format presents a formidable structural and technical challenge. Faculty members who have gone all-in to produce high-quality videos in support of access-enhancing interactive pedagogies now face the massive – and often minimally supported or unsupported – task of producing intelligible transcripts or alternative board notes that meaningfully incorporate mathematical expressions and equations into a screen-readable accessible format. In practice, these well-intended efforts are often hard-pressed to achieve anything approaching genuine utility for the students they are intended to support, and may often fail to achieve even checkbox compliance.

*Corresponding author: looneyc@merrimack.edu

As educators with roughly 300 instructional physics videos between us on our YouTube channels, we faced a challenge that we did not know how to meet. While many excellent tools have been developed to help educators make math more accessible, constraints relating to scope, availability, or overall utility restrict their widespread application to accessible-math video transcription. Mathpix Snip [4] and Purdue University's proprietary ink2html [5] are great tools for transcribing written math or images of math to LaTeX or HTML, but have no audio capability. Equatio [6] can transcribe dictation or convert screenshots to LaTeX or MathML, but cannot do both simultaneously and cannot listen to a video's audio track. The open-source tool AudioTTo [7] utilizes the Faster-Whisper implementation of OpenAI's Whisper [8] to transcribe prerecorded lecture audio into text, which is then fed alongside PDF slides into Google Gemini to generate a structured, LaTeX-formatted summary. VoiceMath [9] is a semi-automatic, human-in-the-loop application designed specifically for video lectures. Developed by the Polin Laboratory at the University of Turin, its availability remains largely restricted to collaborators and institutional affiliates. The web application transcribes the video's audio track while utilizing the Mathpix OCR API [10] to convert visual board note screenshots into LaTeX notation. A human transcriber reviews and edits the resulting output in a specialized workspace to produce structured LaTeX notes. While these are impressive tools, they do not – individually or collectively – achieve what seemingly ought to be possible with generative AI: a free, widely available, mostly automated system that can make integrated use of the audio and visual information in a video, resolve gestural referents, and integrate mathematical formulas and essential visual descriptions directly into a math-accessible transcript.

To address this, we developed a mostly automated transcription workflow using Google Gemini in the AI Studio environment [11] and The LaTeX Project's recent stabilization of automated Structure Element (SE) tagging in LuaLaTeX [12, 13]. The transcribing Gemini AI, guided by our carefully crafted system instructions, listens to the audio track and effectively watches the video track – sampling images at a default rate of one frame per second – of the uploaded or linked video and produces LaTeX-formatted output that includes pedagogically relevant descriptions of visual information. Crucially, all board-written mathematical expressions and equations are transcribed by Gemini into LaTeX and incorporated into the transcription output, which is subsequently pasted into a prewritten LaTeX (`.tex`) shell document with a state-of-the-art LuaLaTeX preamble. After commenting out redundant formatting commands and adding any desired annotations, the resulting document is compiled by a LuaLaTeX engine in a local-machine LaTeX installation to produce a PDF document. Following visual inspection and subsequent validation of PDF/UA-2 and ISO 32005 compliance using PDFix, the accessible PDF transcript can be distributed, posted, or stored with all the flexibility and stability the PDF format affords. All of this is done using freely available tools.

During our late-stage writing of this manuscript, we learned about the YouTube AI Math Transcriber [14], which pairs AI-enhanced audio captions with event-driven video sampling. While the absence of an HTML output option has apparently obscured recognition of its accessible-math potential, in our brief initial tests – carried out on two test videos using the Clean, Reformulated or Examples "Interpretive level" options – the LaTeX output has reliably compiled with our LuaLaTeX preamble to produce math-accessible PDF transcripts with impressive incorporation of board-written or screen-presented equations. While this tool is only available for public or unlisted YouTube videos and requires a paid subscription to transcribe more than 5 videos per month, we believe it merits further investigation.

The rest of this paper is organized as follows. Section 2 describes the free tools used in our transcription workflow, Section 3 presents an overview of our development process, and Section 4 describes the essential products of our development: the AI system instructions that guide transcription and the LuaLaTeX preamble that ensures reliable compilation into accessible PDFs. Section 5 describes the testing of our workflow using 16 videos, and presents a detailed evaluation of the publicly available test transcripts. Section 6 describes our brief experimentation with radically minimalist system instructions, and Section 7 is a guide to the publicly available supplementary materials [15], which include the preamble and system instructions needed to run our workflow and all the cited test transcriptions to allow independent evaluation. Our Conclusion in Section 8 is followed by a clear disclosure of generative AI use.

Before proceeding to Section 2, we note that this manuscript is itself compiled using LuaLaTeX and verified for PDF/UA-2 and ISO 32005 compliance with PDFix Desktop Lite (version 3.2.0) – ensuring that the transcript excerpts presented throughout remain natively math-accessible in PDF format, rather than relying on downstream HTML conversion.

## 2 Overview of Free Tools Used in Our Transcription Workflow

Here we provide an overview of the free tools – Gemini in the AI Studio environment, LaTeX, and PDFix – that we use for transcription, PDF document production, and accessibility checking in our workflow, so that the remainder of this paper can be more easily understood without a technical AI or LaTeX background.

### 2.1 Gemini in the AI Studio environment

There are several factors that make Gemini well-suited for video transcription, especially when used in the AI Studio[1] environment. Gemini has a generous free tier, and it both listens to and watches the uploaded video; by default it samples one frame per second in the AI Studio environment. The combination of audio and visual information helps Gemini build a more accurate representation of the overall presentation, and allows Gemini to include pedagogically essential visual information, including but not limited to presented equations, in its transcription output.

Figure 1 shows a screenshot of the AI Studio environment [11]. A media button at the bottom of the window, selected and shown in the Figure 1 screenshot, allows the user to upload a video or provide a YouTube URL. The blue-outlined YouTube URL feature, which is only available for public videos with creator-allowed embedding, gives Gemini immediate access through a back-end Google connection; this substantially expedites processing by eliminating upload waiting times from the workflow. A box at the top of the right panel allows the user to select the Gemini model. In our tests we used Gemini 3.5 Flash; Gemini 3.7 Flash recently became available on the free tier and should offer higher performance. Right below that is a box for the system instructions, which can be given a name and saved for rapid future redeployment; this is where we paste our detailed transcription instructions. Right below that is a box for the thinking level; we always select “High.” The red-outlined main prompt box, located at the bottom of the window just to the left of the media button, is where we paste a short kickoff prompt that tells Gemini to transcribe the uploaded or linked lightboard, whiteboard, or screencast video according to the loaded system instructions. The main prompt box is also used for any followup dialog with Gemini.

[1] Note on tool selection: while the consumer Gemini chatbot can also transcribe videos, the AI Studio environment currently offers a substantially more generous free-tier video-length upload limit and a more convenient provision for standing transcription instructions.

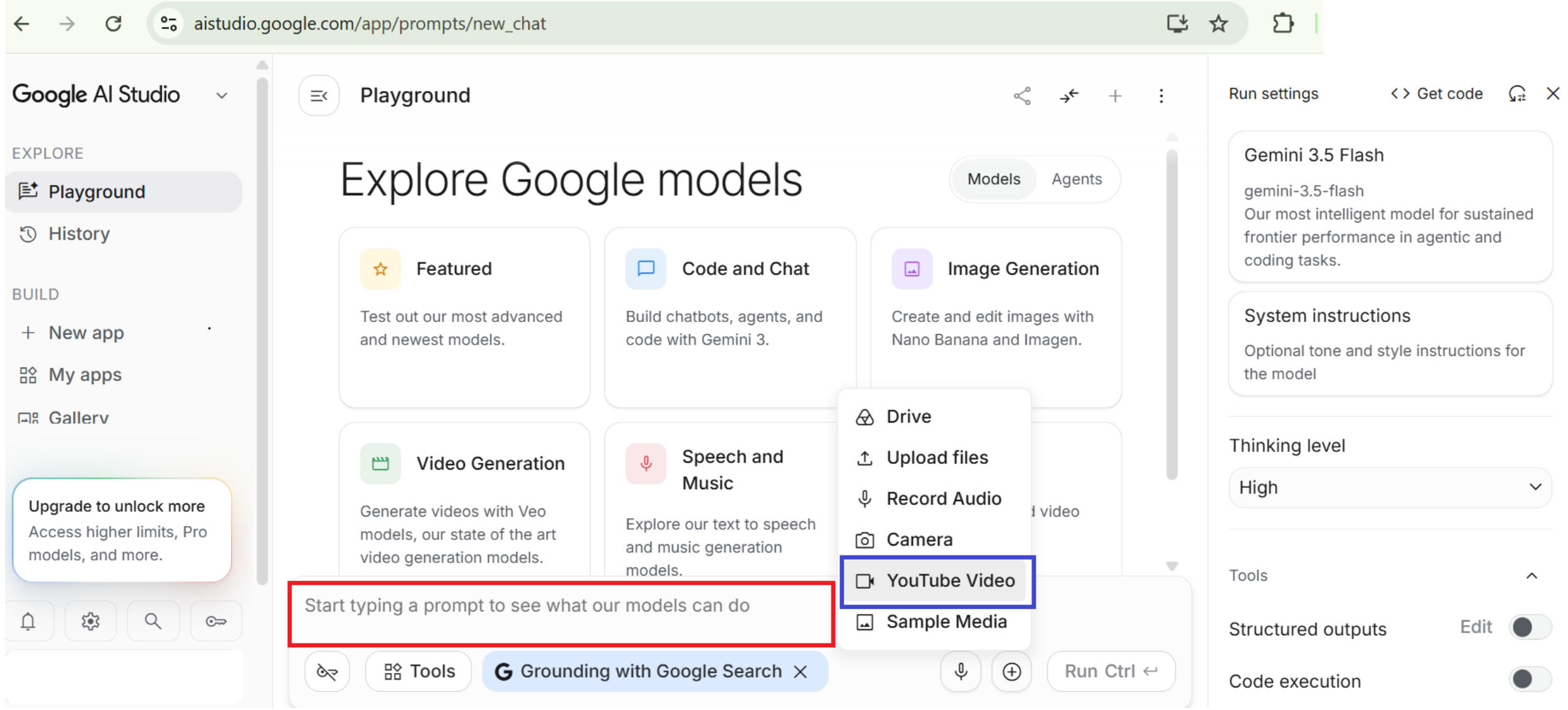

Figure 1: The AI Studio [11] environment.

By far the most time-consuming part of this project was developing, through numerous cycles of testing and revision, system instructions that could consistently deliver professional-quality transcriptions that compiled flawlessly into accessible PDFs. The resulting well-tested v6m system instructions, presented and discussed further below, are available in file `system-instructions-v6m.txt` in the supplementary materials [15]. This file also contains the three different variations of the short kickoff prompt used for lightboard videos, whiteboard videos, and screencast videos, respectively.

Note that while using the free tier of Google AI Studio does not forfeit an instructor's copyright, any data entered may be used by Google for model training and may be reviewed by humans [16]. Consequently, users must strictly avoid inputting protected, sensitive, or proprietary institutional information, and more generally should consult and follow their institution's AI policies.

## 2.2 LaTeX and LuaLaTeX

In very general terms, LaTeX is an advanced typesetting system for mathematical and scientific content. It is not at all like Microsoft Word, which uses special files containing hidden characters and which follows a "what you see is what you get" principle. LaTeX files are simple text files, and LaTeX documents follow a "what you see is what you *mean*" framework. That is, authors make decisions relating to semantic structure – chapters, sections, subsections, inline vs. display math, etc. – and LaTeX takes care of the layout. Authors encode their structural intentions and mathematical representations using commands inserted directly into the document. The to-be-formatted content is preceded within the same file by a preamble that specifies which portions of the vast LaTeX universe are to be brought to bear in the document production process. The file is then handed over to the LaTeX compiler. If everything has been done correctly, the result is a beautiful PDF that has been perfectly formatted in accordance with the preamble and the in-document commands.

It has proven to be incredibly difficult to translate LaTeX's longstanding visual mathematical typesetting supremacy into more generalized mathematical accessibility. Indeed, during LaTeX's ongoing four-decade run as the visual gold-standard for mathematical and scientific typesetting and document production, HTML came into being, progressed through a long and painful math-unfriendly adolescence, and improbably emerged as the gold standard for accessible mathematics, in spite of the fact that HTML relies on web browsers rather than static, universal document viewers. However, since November 2025 it has been possible to have gold-standard mathematical accessibility in a LaTeX-produced PDF through the stable implementation of automated Structure Element (SE) tagging [13], which allows the same semantic and structural information found in accessible HTML plus MathML to be embedded into the PDF tagging tree. Importantly, SE tagging is carried out *automatically* and robustly by the LuaLaTeX compilation engine, provided that the LaTeX has been cleanly written and the document preamble has been correctly structured.

While LaTeX is not known for simplicity or user-friendliness, it is extraordinarily well-suited for an AI transcription workflow; LaTeX is basically code, and AIs excel at coding. Unfortunately, the rigorous LuaLaTeX SE tagging process that makes accessible PDFs possible comes at the cost of substantially increased compilation time, which in turn causes the free tier of web-based Overleaf, the least user-unfriendly LaTeX environment, to time out before even a short 2–3 page LuaLaTeX document can finish compiling. While paid Overleaf tiers, which cost on the order of 15 to 20 dollars per month, can handle much larger jobs without timing out, compilation takes an order of magnitude longer on web-based Overleaf than when using a laptop or desktop LaTeX installation. Consequently, at the time of this writing, we must recommend a local-machine setup, which entails a lengthy and nontrivial download and installation of a TeX distribution (we use TeX Live [17]) followed by a relatively straightforward download and installation of a LaTeX editor. For editing, CWL uses TeXstudio [18] in Windows, and CLD uses GNU Emacs [19] in Linux. Note that Linux users will likely need to bypass their native package managers and perform a manual "vanilla" installation of TeX Live to ensure access to automated SE tagging.

While it is helpful for users of our transcription system to develop a general sense of how LaTeX works, it is not necessary to become a LaTeX coding expert to use our workflow, in which human LaTeX editing normally consists only of copying, pasting, trimming, annotating, and occasional minor manual edits. In our workflow, it is Gemini's job to write LaTeX, and it is better at this job than most humans can ever hope to be.

In the supplementary materials [15], we provide a preamble-shell document (`preamble-shell-v6m.tex`) containing the well-tested v6m LuaLaTeX preamble with activated SE tagging designed to reliably compile with Gemini's LaTeX transcription output to produce accessible PDFs.

### 2.3 PDFix validation test for PDF/UA-2 and ISO 32005 compliance

It is important, when checking a LuaLaTeX-produced PDF for accessibility compliance, to use an up-to-date checking tool approved by the PDF Association [20] because fully compliant PDFs are likely to fail out-of-date tests. We use PDFix Desktop Lite [21], which is straightforward to install, runs natively on multiple platforms, and is free and easy to use for PDF checking. Simply open the PDFix application, load the PDF file to be tested, and then – as shown in Figure 2 – choose **Actions**, then **Validation**, then **Validate with**, and finally **PDFUA-2-ISO32005**. This rigorous test is designed to check for both PDF/UA-2 and ISO 32005 compliance. If the PDF passes the validation check, the phrase “PDFUA-2-ISO32005” will appear in the lower right panel. If it fails, the lower right panel will display the problematic portions of the tagging tree. In the very rare event that a PDF compiles without errors but fails the PDFix validation test, the PDFix-reported issue can be copied and reported back to the transcribing Gemini AI for diagnosis. We do not use PDFix for remediation; we use it only for checking.

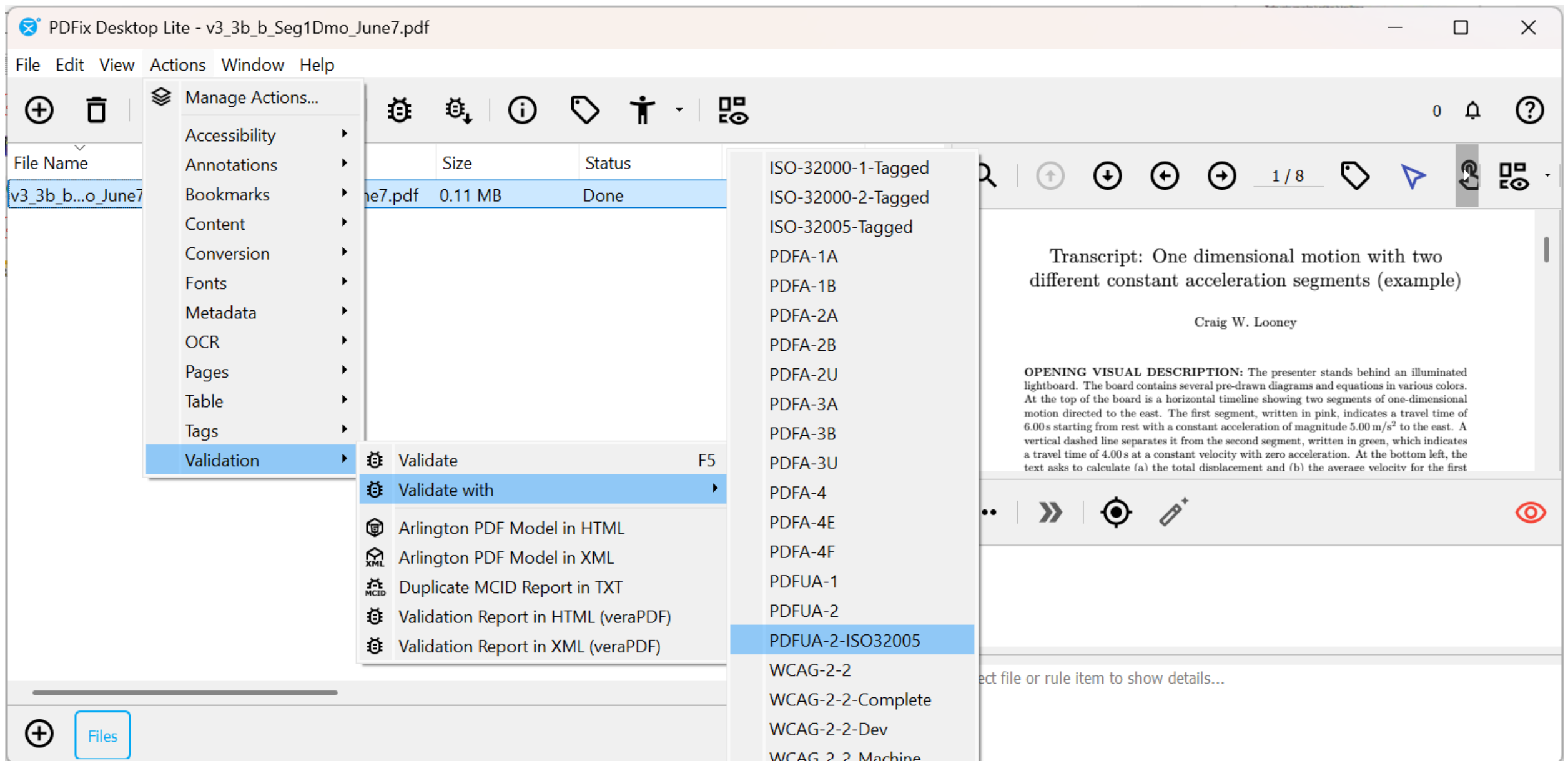

Figure 2: PDFix Desktop Lite [21].

## 3 Development Overview

After we fully committed to clean SE tagging and abandoned reliance on preamble math-coding macros, our LuaLaTeX preamble – which was mostly written and revised by Gemini models under human direction – rapidly assumed its final form. In contrast, the system instructions, which were written with extensive assistance from multiple Gemini consultants, went through more than 50 distinct iterations, including several major reorganizations and at least two complete teardowns and rebuilds. After observing that Gemini-suggested refinements were contributing steadily to overall instruction-set growth, we began to aggressively incorporate anti-bloat measures into our refinement prompts. The aforementioned teardowns and rebuilds involved substantial human-only structural reformulations and wording revisions. While the human-only work was very inefficient from a words-revised-per-hour perspective, it reduced bloat and helped ensure that Gemini was working for us and not the other way around. Important developmental progress resulted from developing LaTeX-compatible adaptations to the traditional transcript structure, abandoning preamble coding macros and relying instead on `amsmath` and other basic packages, leveraging Gemini’s knowledge of math and physics, empowering Gemini to creatively transcribe within carefully considered guardrails, and being willing to blow up and reformulate our subscript framework deep into the project. Since we were continually learning – and continually working to discern best practices from outdated accessibility hacks – throughout the development process, our resulting v6m system instructions might more closely resemble the backward-wired human eye than the logically wired octopus eye. That is: the v6m system instructions are robust and work well, but might contain suboptimal evolutionary artifacts that cannot easily be changed without destroying the well-working system. We will return to this idea later in the paper.

## 4 Description of the v6m Preamble and System Instructions

The v6m preamble and system instructions produced high-quality transcriptions that compiled without error and passed the PDFix PDFUA-2-ISO32005 validation test in 16 out of 17 successive transcription tests involving 16 different test videos. The transcription tests are discussed in the next section; here, we provide a high-level overview of the key features of the preamble and system instructions.

Our v6m preamble requires a stable LuaLaTeX engine and a modern TeX distribution compatible with 2026-standard math-SE tagging. For mathematics and units, the preamble loads only `amsmath`, `unicode-math` (which is automatically loaded by the `newcomputermodern` font package), and `siunitx`. The accompanying system instructions direct Gemini to code all math using only these basic packages, and explicitly warn Gemini that the deprecated `physics` and `physics2` packages, which are not fully compatible with LuaLaTeX, are not loaded and must not be used. AI systems are fully capable of constructing everything from the basic tools afforded by `amsmath`, `unicode-math`, and `siunitx`, and we have found minimalist preambles to be vastly more robust than those that are laden with specialized macros. The only coding macro in the v6m preamble is a fallback `\providecommand` in case Gemini uses the deprecated `\square` command (for example, when transcribing an empty square written by a presenter as a placeholder for a subscript); other such fallbacks can be added as needed to accommodate individualized notation practices.

The complete system instructions consist of a strong expert transcriptionist persona, a short instructions overview designed to give Gemini up-front big-picture context, and a detailed four-part instruction set. The complete system instructions are available in the supplementary resources [15]; we quote the entire persona and overview portions below, and follow this with a general description and discussion of the much longer four-part instruction set.

> PERSONA: Expert AI physics video transcriptionist
>
> You are an expert AI physics transcriptionist converting video lectures into clean, accurate, production-ready LuaLaTeX output that can be combined with a pre-existing state-of-the-art LuaLaTeX preamble and compiled into PDFs that meet PDF/UA-2 (ISO 32005) accessibility standards. Your objective is to ensure that students reading the resulting PDF documents, including blind students using screen-readers, can perfectly reconstruct the spoken equations and the physical, visual arguments made by the presenter.
>
> INSTRUCTIONS OVERVIEW: The instructions below are designed to help you carry out your objective. Part I defines a transcript SEGMENT STRUCTURE based on distinct verbal and descriptive segments; notably, both verbal and descriptive segments can contain math (which does not have its own segment). Part II lays out the GENERAL LaTeX MATH FORMATTING RULES within the segment structure. Part III CHRONOLOGICAL FLOW AND GUIDANCE lays out the transcription process within the segment structure and provides a combination of structured guidance and creative license to empower you to use your AI-enhanced expertise to handle diverse transcription scenarios. Part IV lays out the LuaLaTeX environmental constraints and strict mathematical typesetting rules required to maintain compatibility with the pre-existing LuaLaTeX preamble and the letter and spirit of PDF/UA-2 (ISO 32005) accessibility standards.

—`system-instructions-v6m.txt`, in [15]

Part I of the system instructions, which comes immediately after the instructions overview, defines a "segment structure" that uses boldface labels rather than brackets [ ] to clearly note transitions between verbal transcription and nonverbal descriptions. In early tests using traditional square brackets, several Gemini consultants noted the difficulty of properly closing off display math at the end of a square-bracketed descriptive segment; this specific issue caused numerous compilation failures in early tests. Abandoning square brackets and adopting the segment structure delivered an immediate and substantial improvement in stability and overall performance. We use the word "segment" for these organizational chunks rather than "block" to avoid conflation with other structures that Gemini thinks of as blocks. While our segment structure does not allow brief bracketed insertions of nonverbal description [like this] to be injected directly into the verbal flow, our lightweight segment labels provide a less-ambiguous transition signal for screen reader users, compared to bracketed text.

Part II lays out a "global math rule" to ensure that all LaTeX math is properly wrapped, instructs Gemini how to choose between inline and display math, and lays out further general math formatting rules designed to ensure reliable compilation.

Part III lays out a structured transcription order and fleshes out Gemini's grasp of the segment structure with specific instructions relating to verbal and nonverbal transcription. It also empowers Gemini to insert pointed-to math directly into the verbal flow and to devise alternative ways to describe more complicated interactions with equations, graphs, and diagrams. Part III also instructs Gemini to find accessible ways to describe, for example, board-written "cancel" lines through vanishing terms, and empowers Gemini to correct obvious errors in board-written math and provide a conditional correction report when it does so.

Part IV tells Gemini how to accessibly code undergraduate-level physics and mathematics using the basic tools afforded by the `amsmath`, `unicode-math`, and `siunitx` packages. Importantly, it also lays out carefully constructed accessible subscript transcription rules according to the following scheme:

- All subscript elements consisting of bare numbers, single symbols, or single letters are coded as *math* and typeset snugly together. This ensures attractive visual typesetting while ensuring that all distinct subscript elements (with the exception of adjacent numbers, discussed below) will be read out separately by an up-to-spec screen reader.
- All subscript elements consisting of multi-letter labels are coded as *text*, so that they will be read out as words rather than as individual letters (e.g., "sys" rather than "s y s").
- Subscript elements are normally transcribed in the board-written order. Board-written commas are transcribed as written, but are not artificially injected (e.g., if $v_{x0}$ is written on the board, it is transcribed as $v_{x0}$, *not* as $v_{x,0}$). This approach is reasonably well aligned with Gemini's natural tendencies and accessibly accommodates a wider range of board-written notation compared to schemes that would require Gemini to impose a universal order or insert required comma separators.
- In exceptional cases involving vertically stacked board-written subscripts, the subscript ele-

ments are identified by Gemini and rearranged into an unstacked configuration, with commas or other visual padding inserted as needed. We note that while LaTeX supports stacked subscripts, unstacked subscripts are more robust and are simpler for screen readers to handle.

In general terms, our subscript framework leverages the properties of both LaTeX math and LaTeX text to semantically group and segment subscript content. Unfortunately, directly adjacent bare numbers present a technical semantic separation challenge that cannot be fully solved through SE tagging alone. Consider a hypothetical snippet of board-written mathematics that appears visually as $A_{123}$. There is no way to know, without further context, whether 1, 2, and 3 are separate labels, or whether they together form a single label to represent, for example, a 123-kilogram object, or whether 12 and 3 are separate labels, etc. Note that adjacent bare numbers that are individually wrapped in braces or separated by small math spaces are collapsed by the compiler into a single multi-digit number; consequently, these schemes fail to achieve semantic separation while creating more work for the compiler. While adjacent bare-number subscript elements *can* be cleanly semantically separated through the injection of commas, there are contexts where comma injection would violate accepted notation standards (e.g., Miller indices) or would change the meaning entirely (e.g., tensor subscripts). For this and other reasons – including AI attention management and a desire to preserve board-written notation whenever possible – using comma injection to enforce strict semantic separation was a nonstarter for us in an AI transcription context. It should be noted that modern math-capable screen readers can already be set by users to read out successive numeric subscripts as individual digits, thereby allowing a screen reader user to determine the meaning from context in the same way that a visual reader would. Thus, while adjacent bare numbers are not semantically separated within our coding scheme, our framework is nevertheless generally compatible with a plain-language concept of accessibility. Our purpose in the preceding description is *not* to make a universal assertion about how this or any other accessibility boundary issue ought to be navigated, but rather to provide reasonable disclosure and explanation relating to the treatment of adjacent bare numbers within our subscript framework.

Beyond the general directive to encode all mathematics using only the tools available in preamble-loaded packages (`amsmath`, `unicode-math`, and `siunitx`) and the detailed subscript instructions, the remainder of Part IV is mostly an accessible-math cheat sheet for Gemini to consult as needed to override training data that is shot through with forbidden tool use and accessibly deficient constructions. We do not know how much difference this cheat sheet makes. We think, but are not certain, that the AI-attention cost is low or modest.

## 5 Testing, Evaluation, and Discussion of the v6m System Instructions and Preamble

The following publicly available lightboard [22, 23, 24, 25], whiteboard, and screencast videos were used for testing and evaluation of our v6m preamble and system instructions. Videos 1–10 and 16 were produced by CWL; videos 11–15 were produced by CLD. While videos 10 and 16 are artificial test videos, produced specifically to test and evaluate particular aspects of our transcription system, the rest are real instructional videos. Overall, the videos cover a reasonable sample of mostly introductory physics topics, span three different video types, and were produced by two different faculty members. While this spectrum surely does not encompass the full diversity of the space of introductory physics videos, it represents a reasonably diverse sample.

1. Relative velocity examples (relativistic, 1-D). 15 minutes, 58 seconds. Whiteboard video.

2. Fourier's trick in Dirac notation. 6 minutes, 10 seconds. Whiteboard video.
3. Magnetic field due to 2 long currents (fully vectorized example). 19 minutes, 46 seconds. Lightboard video.
4. Projectile motion example (horizontal travel, launch height, impact velocity vector). 20 minutes, 36 seconds. Lightboard video.
5. Quanton in a box. 9 minutes, 44 seconds. Screencast video.
6. Elevator madness! Using Newton's 2nd and 3rd laws. 14 minutes, 55 seconds. Screencast video.
7. Separation vectors and unit separation vectors. 5 minutes, 16 seconds. Whiteboard video.
8. Taylor series derivation of $v^2 = (v_0)^2 + 2a(x - x_0)$. 7 minutes, 58 seconds. This lightboard video is a supplementary resource for [26].
9. 1-D motion with two different constant acceleration segments (Example). 15 minutes, 35 seconds. Lightboard video.
10. Equationfest, Take 2. 3 minutes, 27 seconds. Lightboard video.
11. J10.1: Introduction to Harmonic Motion. 3 minutes, 24 seconds. Whiteboard video with physical demonstration of an oscillating mass-spring system.
12. Why does a positive phase shift look like a leftward shift of the function? 5 minutes, 18 seconds. Screencast video.
13. Review Problem: Standing Waves. 10 minutes, 49 seconds. Screencast video.
14. Introduction to Gauss's Law 2. 15 minutes, 8 seconds. Lightboard video.
15. Energy in Oscillatory Motion. 18 minutes, 12 seconds. Lightboard video.
16. Conservation of momentum (subscript tests). 8 minutes, 18 seconds. Lightboard video.

### 5.1 Overview of developmental-stage testing

Initial plausibility tests used a short (1 minute, 14 seconds) video not listed above. Early- to middle-stage testing heavily utilized videos 3, 8, and 9. Only after abandoning preamble coding macros and relying entirely on `amsmath`, `unicode-math`, and `siunitx` for all LaTeX math constructions did it become possible to transcribe three different videos in succession without compilation errors. Our 34th set of system instructions (v5b) compiled without errors and passed the PDFix PDFUA-2-ISO32005 validation test in 22 successive transcription tests across 16 videos, including 14 of the 16 videos listed above. Our excitement turned to disappointment upon realizing that our original subscript scheme involving the insertion of "small math spaces" turned out to be yet another outdated accessibility hack that now does more harm than good; a close inspection of the LaTeX also revealed further subscript-related issues that needed to be addressed. After the subscript framework was reformulated as described above, the system instructions progressed through 19 further revision cycles as we worked to stably integrate the new subscript framework and pursue further improvements, ultimately resulting in the v6m system instructions described above and available in full in the supplementary materials [15].

## 5.2 Overview of transcription tests of the v6m system instructions and preamble

The v6m system instructions and accompanying preamble were tested using the 16 videos listed above. All tests were conducted in fresh AI Studio windows using Gemini 3.5 Flash with High thinking level selected. The v6m system instructions were loaded in the system instructions box, a link to the video was provided using the YouTube URL tool shown earlier in Figure 1, the appropriate version of the kickoff prompt was pasted into the prompt box, the prompt window was given a name for future reference, and the Run button was clicked. The entirety of Gemini's transcription output, including any Gemini-produced correction logs appended as comments, was copied and pasted into a LaTeX shell document containing the v6m preamble at the beginning and a Transcription Information section at the end. Redundant `\begin{document}` and `\end{document}` commands were commented out prior to compilation. Any correction logs were preserved in their original output positions within the LaTeX source files and also copied and pasted, with line breaks inserted, into verbatim environments within the Transcription Information sections so that the correction logs could be made safely visible in the resulting test PDFs. Each compiled PDF was tested for PDF/UA-2 and ISO 32005 compliance using PDFix Desktop Lite.[2] After an initial read-through, transcription notes were added to the Transcription Information section and the document was recompiled and re-checked for accessibility compliance. No edits were made within the body of any test transcripts.

The single-shot transcriptions of videos 1–8 and 10–16 compiled without errors and passed the PDFix PDFUA-2-ISO32005 validation test. Video 9 required two transcription attempts. The first LaTeX transcription of video 9 had an inline math expression that was not properly wrapped with dollar-sign delimiters; consequently the resulting PDF compiled with errors and had 4 lines of run-together text that was wrongly coded as math. The second transcription of video 9, undertaken in a fresh AI Studio window, compiled without errors and passed the PDFix PDFUA-2-ISO32005 validation test.

The LaTeX source files and all compiled PDFs produced using the v6m system instructions and preamble are available in folders "a-latex-files-v6m" and "b-pdf-files-v6m" in the supplementary materials [15]. Note that within the LaTeX source files, the preamble is referred to generically as "version 6x" rather than v6m. However, the preamble code has been unchanged since the early v5-series tests (v5a). We began using the generic "version 6x" name so that we could focus on testing each new system instructions iteration without having to worry about continually updating the name of an unchanging preamble in the comments at the top of the shell file.

Below, we discuss and evaluate – in considerable detail – important transcription aspects beyond reliable compilation and PDF/UA-2 and ISO 32005 compliance. However, some – perhaps many – readers might prefer at this point to read through one or more v6m test transcripts in the supplementary materials [15] before deciding whether they want to read, skim, or skip entirely our detailed discussion. Nevertheless, those who might use our system should read Subsection 5.10, which discusses transcript inspection, cleaning, and finalization.

[2] PDFix Desktop Lite version 3.0.0 was used for all date-of-compilation checks noted in the Transcription Information section of each test transcript in the supplementary materials [15]. All test transcripts were checked again on August 18, 2026 using version 3.2.0. All test transcripts that passed the version 3.0.0 compliance check also passed the version 3.2.0 check.

### 5.3 Accuracy of transcribed audio

Since we have empowered Gemini to replace ambiguous "this equation" statements with pointed-to math and have disallowed the injection of bracketed nonverbal insertions [like this] into verbal transcription, and since our instructions prioritize pedagogical coherence and flow over strict verbatim, we have not undertaken exhaustive word-by-word comparisons of the transcripts with the source videos. Instead, our general approach has been to read through the transcripts to see if they make sense, keeping an eye out for possible errors that might cause confusion. With the exception of some ambiguous references (e.g., "this", "here", "this equation") that Gemini fails to replace with a LaTeX rendering of pointed-to math or describe in a descriptive **ACTION** segment, the verbal transcriptions are generally clear. Indeed, in spite of the latitude we have given Gemini, the strong "expert AI physics transcriptionist" persona apparently drives Gemini toward lightly cleaned verbatim whenever possible. Importantly, Gemini brings a strong mastery of physics and mathematics to transcription. An examination of Gemini's "thoughts" while transcribing indicates that it works through the logic of each video presentation; an example of Gemini's transcription thoughts is available in the supplementary materials [15] and excerpts from those thoughts are discussed further below. It should be noted that the context provided by the visual information in each presentation, including but not limited to written equations, informs Gemini's understanding of the audio track. Consequently, if a presenter mumbles the word "force," it is unlikely that Gemini would ever transcribe it as "forks." In reading through the transcripts, we believe the verbal accuracy – as it pertains to understanding the presented content – is very good, and that a professional transcriber with a degree in physics would be hard-pressed to do better.

### 5.4 Accuracy of transcribed mathematics

Overall, we have found the accuracy of transcribed mathematics to be very good. As noted above, Gemini does not mindlessly copy letters, symbols, shapes, etc.; rather, it thinks through what it is transcribing. Gemini's overall grasp of the lecture can help it correctly interpret things that might have been written too sloppily or too small to unambiguously identify in isolation. We have found that Gemini will often correct obvious board-written physics or math errors on its own. The error correction and reporting instructions (`system-instructions-v6m.txt`, III.8, in [15]) were designed to positively leverage this tendency while providing guardrails to prevent Gemini from hallucinating errors in order to fix them. We believe that well-defined error-correction empowerment also helps reduce potential conflicts between the persona-driven goal of transcription fidelity and the vast math and physics training data that conditions Gemini to expect correctly presented math and physics. For examples of correction logs, see the v6m transcripts for videos 1, 3, 6, 8, 9 (test 2), 10, and 13 in the supplementary materials [15]. The correction logs, which are appended to Gemini's LaTeX output as comments, would normally be visible only in the LaTeX source files. In order to make the correction logs safely visible in our test PDFs, we have copied and pasted each correction log into a verbatim field in the Transcription Information section at the end of each document and inserted manual line breaks.

While to a first approximation we have observed that Gemini's understanding of the presentation gives it a capacity – similar to and sometimes even superior to that of a human physics-trained transcription expert – to identify and correct errors in presented math, we have also observed cases where Gemini falls short of this standard. For example, in the v6m transcription of video 1, Gemini consistently used an uppercase $V$ rather than a lowercase $v$ for velocity, and in the v6m transcriptions of videos 4 and 8 Gemini used a combination of uppercase $V$ and lowercase $v$ to represent velocity. While this could easily be addressed in a real-world transcription scenario by asking Gemini to go back through the transcript and use only lowercase $v$ to represent velocity, a human physics-trained expert transcriber would never have used an upper-case $V$ to represent velocity in these videos. Interestingly, Gemini's conflation of $V$ and $v$ happened in spite of the following directive in the v6m system instructions:

> CONTEXTUAL CASE CONSISTENCY: For scale-dependent variables that look identical in handwriting except for size (e.g., v/V, x/X, p/P), use the physical and lecture context to resolve visual ambiguities. Maintain absolute casing consistency for each distinct physical quantity across the entire transcription, avoiding arbitrary case transitions within the same derivation.
>
> —`system-instructions-v6m.txt` (II.e), in [15]

The incorporation of this instruction during the v6-series testing cycles made a noticeable improvement compared to earlier sets of system instructions. Yet in spite of this clear v6m directive, and in spite of Gemini's demonstrated ability to utilize presentation context and its knowledge of math and physics to correct errors and enforce other aspects of consistency, Gemini still struggles with particular case-consistency issues that would be easily and unconsciously handled correctly every time by an expert human transcriber.

Gemini also occasionally exhibits other interesting nonhuman transcription behavior that potential users of our system should be aware of. In the single v6m transcription of video 3, Gemini systematically reversed the order of board-written subscripts from $\hat{r}_{1\perp}$ to $\hat{r}_{\perp 1}$ and from $\hat{r}_{2\perp}$ to $\hat{r}_{\perp 2}$; since the reversal was systematic, the transcript itself is completely clear and consistent. Subscript-order reversal has been very rare in our tests. This particular video has been transcribed at least ten times by various iterations of the system instructions, but the subscript order was only reversed one other time.

In one earlier-version transcription of video 3, Gemini transcribed every single board-written instance of $\hat{x}$, $\hat{y}$, and $\hat{z}$, respectively, as $\hat{\imath}$, $\hat{\jmath}$, and $\hat{k}$. This substitution was done with complete consistency. This happened in only one test transcription, and only with this particular video. Video 7 also heavily uses $\hat{x}$, $\hat{y}$, and $\hat{z}$, but Gemini never changed these to $\hat{\imath}$, $\hat{\jmath}$, and $\hat{k}$, although it should be noted that video 7 has had far fewer test transcriptions than video 3. Similarly, in the v6m transcription of video 10, Gemini transcribed the board-written volume element $d\tau$ as $dV$. Versions of this video have been transcribed several times by multiple iterations of the system instructions, and Gemini has made this exact same substitution every single time, presumably due to the dominance of $dV$ in Gemini's training data. While these notation changes do not impact the intelligibility or coherence of these particular transcripts, users of our system should be aware of the possibility of such notation changes.

Finally, we note that while the v6m transcription of video 10 – which featured 17 pre-written equations grouped under five headings – is quite good overall, earlier versions of this video with less anchoring context caused Gemini, directed by earlier sets of system instructions, to make more transcription errors and to occasionally hallucinate equations that it presumably expected to see. However, Gemini normally has no problem accurately transcribing large numbers of pre-written equations – for example, in the "Opening Visual Description" segment at the beginning of each transcript – because in a normal video, there is plenty of subsequent presentation context. Importantly, Gemini does not transcribe as it watches and listens; rather, it watches and listens and then transcribes.

While potential users should be aware of these occasionally occurring issues – all of which could easily be fixed by Gemini upon request (e.g., "Gemini, in some places in the transcript you used capital V for velocity; please correct the transcript to use only lowercase v for velocity") – it is important to maintain reasonable perspective. We think that a professional human transcriber with a degree in physics would struggle to match the overall mathematical accuracy of the v6m transcriptions. While a human transcriber would never use an uppercase $V$ for velocity, we believe a human transcriber would be substantially more likely to make random or mundane transcription errors. Furthermore, in an accessibility context, we need to factor in coding accuracy. Few professional transcribers are likely to match a well-instructed Gemini model with regard to cleanly coding accessible mathematics in LaTeX or HTML.

### 5.5 Subscripts

While as noted above we have observed that Gemini will very occasionally reverse the order of board-written subscript elements, the v6m subscript instructions robustly handled a wide range of board-written multiple-subscript configurations. Here are some representative examples involving multi-element subscripts. These and all subsequent v6m LaTeX coding examples and passage excerpts have been copied and pasted from the v6m transcription LaTeX source files in the supplementary materials [15]; the source video is noted for each presented example throughout the remainder of this paper. Board-written stacked subscripts will be discussed further below.

- Video 1: $V_{MS} = \frac{V_{M\square} + V_{\square S}}{1 + \frac{V_{M\square} V_{\square S}}{c^2}}$. While as noted above, Gemini used uppercase $V$ for the velocity, all the subscripts in this video were correctly transcribed. Importantly, the single-letter relative velocity subscripts (e.g., $M$ for missile, $S$ for ship) were correctly coded as math rather than as text so that they can be read out individually by a screen reader. The board-written square was coded using the deprecated `\square` command, but was converted to the correct coding by the earlier-noted preamble `\providecommand` macro.
- Video 2: $\langle \psi_m \mid \psi_n \rangle = \delta_{m,n}$. Note that the comma was board-written, not artificially injected.
- Video 3: $\vec{B}_1 = \frac{\mu_0 I_1}{2\pi r_{\perp 1}} \left( \hat{I}_1 \times \hat{r}_{\perp 1} \right)$. As noted earlier in the paper, the order of subscript elements in unit vectors such as $\hat{r}_{\perp 1}$ was systematically reversed by Gemini. Otherwise the subscripts were well-coded.
- Video 6: $F_{A \text{ on } B} = F_{B \text{ on } A}$. Inspection of the LaTeX source file reveals that $A$ and $B$ were correctly coded as math, while " on " was coded as text with padding spaces on either side in the text field.
- Video 6: $\vec{F}_{\text{fl on } 30} = (390\,\mathrm{N}, \uparrow)$. Inspection of the LaTeX source file reveals that the number 30 was correctly coded as math and "fl on " was correctly coded as text with appropriate padding spaces inserted. Note that "fl" has been text-coded because it is a multi-character abbreviation for floor.
- Video 7: $\hat{r}_{1\to 2} = \left( \frac{\Delta x}{r_{1\to 2}} \right) \hat{x} + \left( \frac{\Delta y}{r_{1\to 2}} \right) \hat{y} + \left( \frac{\Delta z}{r_{1\to 2}} \right) \hat{z}$. Gemini coded this perfectly.
- Video 16: $p_{xi,\text{sys}} = p_{xf,\text{sys}}$. The board-written comma was correctly transcribed. "sys" was correctly coded as text, and everything else was correctly coded as math.

In our inspections of all 17 v6m test PDFs and source files, we have not found any cases – except for the video 3 subscript-element order reversal noted earlier – where Gemini did not follow the ordinary (unstacked) subscript rules, although it is possible that there are a small number of deviations that have escaped our scrutiny.

As noted earlier, for both stability and accessibility reasons, the v6m system instructions direct Gemini to unstack board-written stacked subscripts. Videos 10 and 16 both contain board-written stacked subscripts; these were correctly unstacked in the v6m transcriptions. However, it should be noted that in these videos, which were created specifically for testing our transcription system, the presenter used the words "stacked subscripts" when pointing to the equations containing stacked subscripts. This in-video context was provided to help counter the otherwise artificially difficult "equation sheet" aspect of these videos, but it likely gave Gemini an unfair advantage.

In video 9, equations for the $x$-component of the average velocity and average acceleration were board-written in a stacked configuration, with "$x$" above "avg," in order to preserve board space. As noted previously, there were two v6m transcriptions of video 9, because the first transcription had a compilation error. In both v6m transcriptions, Gemini correctly identified and coded "$x$" as math and "avg" as text. In the first transcription, the unpacking order was exactly as specified in the instructions, while in the second transcription, the order was reversed, with the bottom subscript appearing first rather than last in the unstacked configuration, but at least the subscript information was accessibly preserved. These tests indicate that the unstacking rules, while likely less reliable than the rest of our subscript system, should at least give Gemini a fighting chance to preserve the information in board-written stacked subscripts in real instructional videos.

Overall, our v6m subscript system has enabled Gemini to robustly and accessibly transcribe a reasonably wide range of board-written subscript configurations. This performance comes at the cost of sustained attention demand on the Gemini transcriber. The absence of any sign of reduced overall transcription quality in the longer videos, with the possible exception of the compilation error in the first transcription of video 9, indicates that the attention burden is manageable for videos up to 20 minutes in length.

### 5.6 Mathematics beyond introductory calculus-based physics

While the reliable transcription of introductory physics videos has been the primary focus of our development, we have taken care to avoid unnecessarily limiting the applicability of our transcription system. The extended accessible-coding cheat sheet in Section IV of the system instructions addresses most of the mathematics used in a typical undergraduate physics curriculum. While the cheat sheet does not cover the entirety of undergraduate mathematics, the requirement to code all mathematics using the basic tools of `amsmath` and `unicode-math`, rather than using less-general constructions from physics-specific packages, greatly enhances the probability that Gemini will correctly transcribe board-written mathematics regardless of physical application.

While our testing of higher-level math and physics transcription has been limited, the successful v6m transcriptions of videos 3, 5, 10, and 14 are indicative of generalized capability. Video 3, which was developed for an undergraduate quantum mechanics course, extensively uses Dirac notation. Video 5, developed for an intermediate-level "modern physics" course, derives the energy eigenstates of the one-dimensional infinite square well. Video 10 includes the Maxwell equations in differential form, the one-dimensional and three-dimensional Schrödinger equations, the divergence theorem, Stokes' theorem, a triple integral, and the summation form of the Taylor series expansion of $f(x)$, which Gemini corrected to begin with $n = 0$. Video 14, while made for an introductory physics course, presents the integral form of Gauss's law; impressively, Gemini corrected a subtle error in the board-written notation relating to the correct technical use of closed-integral notation.

The transcription system should reliably handle the mathematics of core engineering courses such as statics, dynamics, and fluid mechanics. However, it should be noted that while the v6m system instructions make special provisions for coding SI units, they contain no special instructions for English units.

## 5.7 Incorporation of board-written and pointed-to math into verbal and descriptive segments

The v6m system instructions reliably ensure that all board-written mathematics is accessibly incorporated into a verbal segment or a descriptive segment or both. Generally speaking, the v6m system instructions (III.5) direct Gemini to inject LaTeX transcriptions of pointed-to math in place of – or immediately following – ambiguous verbal references (e.g., "this equation") whenever possible, and to use nonverbal **ACTION** segments to describe more complicated equation interactions. For a combination of pragmatic, technical, and accessibility reasons, the v6m instructions expressly forbid the normally conventional injection of square-bracketed visual descriptions into the verbal flow; to somewhat oversimplify, square brackets greatly increase the probability of compilation errors due to clashes with LaTeX math and provide a less clear transition signal for screen reader users than our lightweight segment labels. The resulting binary aspect of these instructions has been partially but not completely smoothed out over many iterations of testing and refinement.

The following passage from the v6m transcription of video 8 contains a very well-executed sequence of LaTeX injections into the verbal flow, followed by a clear visual description of a more complicated interaction.

> **PRESENTER:** The second derivative of $v^2$ with respect to $x$ will just be the derivative with respect to $x$ of the derivative of $v^2$ with respect to $x$:
>
> $$\frac{d^2(v^2)}{dx^2} = \frac{d}{dx}\left(\frac{d(v^2)}{dx}\right)$$
>
> Now, the derivative of $v^2$ with respect to $x$, we've already worked through that, that's $2a$:
>
> $$\frac{d^2(v^2)}{dx^2} = \frac{d}{dx}(2a)$$
>
> Now, if the acceleration is constant for an interval of motion, that means it's constant with respect to both time $t$ and position $x$. So, if the acceleration is constant, then the derivative with respect to $x$ of the acceleration is equal to 0 if the acceleration is constant:
>
> $$\frac{d^2(v^2)}{dx^2} = 0 \quad (\text{if } a = \text{const.})$$
>
> So in the case of constant acceleration, this quantity is 0, and that kills off this term of the sum.
>
> I need to emphasize that by 0, I don't merely mean momentarily 0. I mean 0 and constant! That is, in the case of constant acceleration, the second derivative with respect to $x$ of the velocity squared is 0 and constant.
>
> **ACTION:** The presenter writes "ZERO & CONSTANT!" and "if a = const." in green chalk directly under the second derivative bracket in the main Taylor expansion.

—Video 8

The v6m video 7 transcript contains an impressive sequence of LaTeX injections into the verbal flow of an unbroken presenter segment running from the bottom of page 1 to the top of page 3; here is an excerpt:

> Alright, so, basically this is our three-dimensional vector representation of that separation vector:
>
> $$\vec{r}_{1\to 2} = (\Delta x)\hat{x} + (\Delta y)\hat{y} + (\Delta z)\hat{z}$$
>
> From there we can get the magnitude of that separation vector. We can represent it this way, $r_{1\to 2}$, or this way, $|\vec{r}_{1\to 2}|$, and the magnitude is constructed basically using the three-dimensional Pythagorean theorem—that is, the square root of the sum of the squares of the appropriate components:
>
> $$r_{1\to 2} = |\vec{r}_{1\to 2}| = \sqrt{(\Delta x)^2 + (\Delta y)^2 + (\Delta z)^2}$$
>
> Alright, once we have those things, we can get a unit vector. That's just the directional part of the separation vector, and for many reasons, this is an important vector to construct as well. So, when we're constructing the unit vector corresponding to the separation vector, we use for the symbol a little caret on the top, $\hat{r}_{1\to 2}$, to represent the unit vector. And we can construct the unit vector by taking the original vector and dividing by its corresponding magnitude:
>
> $$\hat{r}_{1\to 2} = \frac{\vec{r}_{1\to 2}}{r_{1\to 2}}$$
>
> Running that through, if we want to look at the components of the resulting unit vector, those components will necessarily be as follows. And then this is the three-dimensional representation of that unit vector corresponding to the separation vector from Location 1 to Location 2:
>
> $$\hat{r}_{1\to 2} = \left(\frac{\Delta x}{r_{1\to 2}}\right)\hat{x} + \left(\frac{\Delta y}{r_{1\to 2}}\right)\hat{y} + \left(\frac{\Delta z}{r_{1\to 2}}\right)\hat{z}$$
>
> —Video 7

We note that the v6m transcripts also contain ambiguous words or phrases (e.g., "this", "this equation") that have not been replaced by LaTeX math. There are many reasons for this. The gesture might have been too short for Gemini to resolve at the default sampling rate of one frame per second. The gesture itself might have been ambiguous, or there might not even have been an accompanying gesture. Importantly, we have not found any examples in the v6m transcriptions in which Gemini has made things worse by guessing or hallucinating incorrect LaTeX injections into the verbal flow.

While the v6m transcripts feature many examples of well-executed LaTeX math injection, Gemini also frequently utilizes alternating verbal **PRESENTER** and descriptive **ACTION** segments to incorporate interactions with board-written mathematics into the transcript. Here is a well-executed example from the v6m video 14 transcript:

**ACTION:** On the right side of the board, the presenter writes the substitution of the fields and differentials in white:

$$\oint \vec{E} \cdot d\vec{A} = \int (E_0 \hat{z}) \cdot (\hat{z}\, dA) + \int (E_0(-\hat{z})) \cdot (-\hat{z}\, dA)$$

**PRESENTER:** Okay, so that's that. Looks kind of messy, but we just have to handle the unit vector stuff, right? So, $\hat{z} \cdot \hat{z}$, that's 1, no problem. $-\hat{z} \cdot -\hat{z}$—okay, that might be confusing, but the two negative signs cancel, the $\hat{z} \cdot \hat{z}$ then is 1 again. So actually, both of those are 1, and we get a nice expression, getting rid of the vectors out of this integral.

**ACTION:** The presenter erases the vector dot product equation on the right and writes in white:

$$\oint \vec{E} \cdot d\vec{A} = \int E_0(z)\, dA + \int E_0(z)\, dA$$

He then writes $dx\, dy$ in orange underneath each $dA$ term.

—Video 14

While it would be possible – and, we think, defensible – to further enhance pedagogical flow by removing the descriptions in the final **ACTION** segment and dropping the final display equation into the verbal **PRESENTER** segment as a generalized extension of pointed-to math, the alternating segment transcription quoted above flows reasonably well and preserves more flavor-of-presentation information. We are happy with the current balance between math injection and segment alternation, and we would also be happy with a slightly different balance in either direction.

While it is not difficult to find unimpressive examples of mathematical integration in the v6m transcripts, there are inherent challenges in balancing flow and literal fidelity when translating multi-channel simultaneous or rapidly alternating information into an accessible single-channel transcript. Our well-motivated decision to forbid bracketed visual descriptions within the verbal flow increases those challenges. Consequently, our system sometimes delivers "adequate but choppy" or mediocre mathematical integration. But more often than not, the integration of LaTeX math ranges from quite good to very good, and sometimes it is spectacular.

## 5.8 Visual descriptions beyond LaTeX math

With regard to describing information in – and presenter interactions with – diagrams and graphs, the performance of our system is not as strong as in the previously discussed evaluation areas. For example, in the 2nd half of the v6m transcription of video 15, the presenter's interactions with graphs are difficult to fully understand from the transcript alone.

We think that the performance-limiting factor is not Gemini itself, but rather the inherently fuzzier and more qualitative nature of graphical and diagrammatic information and interactions, which in turn makes it difficult to write and attempt to finely tune instructions that reliably bring forth Gemini's full capabilities. A board-written equation contains information that can be precisely transcribed. In contrast, when a presenter points to a general location on a graph or draws an arrow on a diagram, the description must generally be qualitative unless there is independent unambiguous information. Is the arrow drawn down and to the right? How far down, and how far to the right? Is the board-written direction a rough sketch, so that a careful literal description would be misleading? What region of the graph did the instructor point to? I think they pointed to the region between $x = 2$ and $x = 4$, but would $x > 0$ be a more pedagogically appropriate description?

It is hard enough for a human to calibrate appropriate descriptions in specific cases, let alone write instructions for an AI that will reliably generate excellent results. Thus, the best we have been able to do with regard to information in and interactions with graphs and diagrams is to provide reasonably thoughtful general instructions and hope for the best. Not surprisingly, the results have been somewhat uneven. However, it is not difficult to find examples of reasonably good performance in the v6m transcriptions, as the following excerpts illustrate:

> **ACTION:** The presenter draws a yellow vector arrow pointing horizontally to the right from $(0, -b, 0)$ and labels it $\vec{B}_1$. He draws a square right-angle symbol between the $y$-axis and the vector $\vec{B}_1$. (Video 3)
>
> **ACTION:** On the left side of the diagram, the presenter draws a vertical pink dashed line down from the launch point to the level ground, labeling it $\Delta y$. (Video 4)
>
> **ACTION:** A hand-drawn free-body diagram of the 20-kg block appears on the left side of the screen. A square representing the block is labeled "20 kg". An arrow labeled $F_{30 \text{ on } 20}$ points upward from the top of the block. An arrow labeled "196 N" points downward from the bottom of the block. A vertical coordinate axis on the far left has a blue arrow pointing upward labeled $y$. (Video 6)
>
> **ACTION:** The presenter updates the query to "plot sin(x+1) from -10 to 10". The resulting plot shows the sine wave shifted to the left, with its zero-crossing at $x = -1$. The presenter then changes the query to "plot sin(x+2) from -10 to 10". (Video 12, describing an interaction with Wolfram|Alpha).
>
> **ACTION:** Within the pipe, the presenter draws a single quarter-wavelength curve starting at a node on the left wall and opening to a wide antinode at the right opening, then draws its mirror image to complete the envelope. Below this pipe, he draws two more identical pipes to represent the next two higher modes. (Video 13)
>
> **ACTION:** Using an orange marker, the presenter draws a three-dimensional rectangular prism (a Gaussian box) that pierces through the plate. The ends of the prism are squares of area $A$. (Video 14)
>
> **ACTION:** On the right side of the board, the presenter draws a second set of axes, with vertical axis labeled $E$ and horizontal axis labeled $x$. He then draws a curve resembling a cubic function $y = x^3$, passing through the origin. (Video 15)

The v6m transcription of video 11, which includes a physical demonstration involving a real mass-spring system, contains an easy-to-miss but spectacular example of Gemini's potential with regard to visual perception and description. About 2 minutes and 20 seconds into the video, the presenter hangs a 0.5 kg mass on a spring, but refers verbally to the mass as 5 kg in the audio track. Gemini seamlessly transcribed the physically correct 0.5 kg mass rather than the actually spoken 5 kg mass into both the verbal flow and subsequent visual description:

> **PRESENTER:** So the key model that we're going to use to study harmonic motion is a mass on a spring. So I have a spring here hanging vertically from a metal bar. I'm going to hang a 0.5 kg mass off of it.
>
> **ACTION:** The presenter hangs a cylinder of mass $m = 0.5$ kg onto the bottom hook of

> the spring. The spring stretches downward until the mass comes to rest at its equilibrium position.
>
> —Video 11

Incredibly, Gemini correctly executed this substitution using only the visual appearance of the mass in the video; the physically correct mass value is not spoken or written anywhere in the video. The following excerpts from Gemini's thoughts while preparing the transcription indicate it gave careful consideration to correctly transcribing the mass value. Note that in these excerpts, Gemini's natively output typewriter-style quotation marks and apostrophes have been converted to their curly unicode counterparts.

> I am now focusing on accurately transcribing the lecture's initial setup and the presenter's description of a spring-mass system. I'm carefully scrutinizing the audio at 2:22, grappling with the mass value. The goal is to obtain the exact number the presenter uses for the mass. Is it 5 kg, 0.5 kg, or something else entirely? The transcript must be accurate!
>
> … I've confirmed the mass being used is 0.5 kg, despite a faint audio utterance of "point five". Based on the context of the experiment and the equipment, 0.5 kg is correct. I am meticulously reviewing the siunitx package rules to adhere to standard LaTeX formatting and will use 0.5 kg within math mode. I'm focusing on ensuring proper LaTeX syntax for all quantities and units.
>
> —excerpts from Gemini's thoughts while transcribing video 11 using the v6m system instructions (`z-gemini-thoughts-while-transcribing-video-11-v6m.txt`) [15]

We have carefully listened to the spoken value several times. The presenter does *not* utter "point 5," faintly or otherwise. We must conclude that factors such as the visual appearance of the mass and spring and the visual appearance of the mass-spring oscillations were so much more consistent with training-data expectations for a 0.5 kg mass than for a 5 kg mass that Gemini imagined that it "heard" a "point 5" that was not actually spoken.

The lesson here is not to be terrified by a minor hallucination. Humans regularly make much larger context-driven perception errors. Rather, the lesson is that Gemini has an impressive capacity for visual perception and inference that could presumably be leveraged by a well-engineered set of instructions to produce outstanding visual descriptions of non-equation visual information and interactions.

### 5.9 Evaluation and discussion of the Opening Visual Description segments

While the Opening Visual Description segments contain elements and aspects that have already been discussed, their purpose – and an evaluation in the context of that purpose – warrants further discussion. In early testing, we found that, without any special prompting, Gemini models almost always precede verbal transcription with a description of the initial appearance of the board or screen. Eventually, we wrote specific instructions for a special Opening Visual Description segment to enhance descriptive quality and consistency and to help anchor, at the outset, a higher level of descriptive performance throughout the transcript. While the Opening Visual Description segments still contain a fair amount of variability with regard to inline vs. display math formatting decisions, the overall descriptions are consistently thorough and intelligible, and we believe that the anchoring function has indeed been helpful.

However, while the Opening Visual Description segments are working as intended, we think there is likely to be a wide range of opinion regarding how much detail such segments should contain. Our well-motivated design decisions, which are now deeply baked into our instruction set and cannot be undone without negatively impacting overall performance, have driven the level of detail to the high end of the spectrum. It should be noted, however, that the first inserted section heading in each transcript immediately follows the Opening Visual Description. With advance notice about this feature, students using screen readers in video-intensive courses could skip directly to the first heading if they have found previous Opening Visual Descriptions to be unhelpful.

### 5.10 Transcript cleaning and finalization

While the overall structural quality and mathematical accessibility of the unedited v6m transcripts are impressive, the unedited transcripts created through our controlled one-shot testing process often contain minor defects that should be corrected prior to transcript finalization. As noted above, case errors or inconsistencies (e.g., $V$ instead of $v$ for velocity) are common. Gemini cannot be expected to correctly distinguish between similar-sounding names with different spellings (e.g., Jane, Jayne, Jain, Jaine). In our many rounds of testing, we have found – in the absence of spoken presenter self-reference – occasional injections of hallucinated names into the opening visual description segments (e.g., "the presenter, Chad"). The opening visual descriptions may also contain unnecessary details relating to presenter appearance. We therefore suggest the following real-world workflow adaptations to obtain maximal cleaning per unit of effort spent.

First, regardless of whatever else might be done, we recommend that the transcript be read through from beginning to end to check for reasonable coherence and to identify obvious errors or defects that must be addressed. Some users with extensive LaTeX experience might prefer to read the LaTeX file, but most people should carry out an initial compilation and read through the resulting PDF. Isolated errors such as localized misspellings or obvious presentation mistakes left uncorrected by Gemini (e.g., presenter said "constant velocity" but clearly should have said "constant acceleration") are most easily corrected by hand in the LaTeX file. Other errors, such as systematically incorrect use of uppercase $V$ for velocity, could also be hand corrected, but can more easily and reliably be fixed by instructing Gemini with a followup prompt: "… please go through the transcript and make sure that lowercase v is always used for velocity, velocity components, and speed."

Some systematic shortcomings might be easily and efficiently addressed through specific additions to the kickoff prompt. For example, if a presenter's name is always misspelled, a brief instruction about the correct spelling could be added. If Gemini often uses $V$ instead of $v$ for velocity, a targeted kickoff-prompt instruction to always use lowercase v for velocity, added only to the kickoff prompts for videos involving velocity or speed, might be helpful. In order to avoid kickoff-prompt bloat, we recommend restraint with this device.

In the event of compilation errors, which should hopefully be rare, the error reports can be copied and reported to Gemini by pasting them into the bottom prompt box, preceded by something like the following: "There were some compilation errors. Don't attempt to re-transcribe just yet. Instead, can you first try to determine what might have gone wrong, and then explain it to me, from the following error report?" Our experience from the error-laden development process is that Gemini can almost always diagnose, explain, and correct the underlying error or errors if it is given the opportunity to think before it attempts to re-transcribe. Thus, while it is of course helpful to develop a very general overview-level grasp of what LaTeX is all about, it should not normally be necessary for users of our workflow to hand-code LaTeX mathematical formatting. In our transcription workflow, that is Gemini's job. Note that if the video is especially long, it might be necessary to wait a few minutes before re-prompting Gemini, in order to avoid going over the free-tier output rate limit. (This does not trigger charges; rather, it temporarily shuts off access to AI Studio, sometimes until midnight.)

After correction of any discovered errors and subsequent recompilation, the new PDF should be reread, further corrections made if necessary, and the process repeated until a clean reading results. We plan to post each finalized PDF in a publicly available but privately controlled Google Drive folder, and place a link to the transcript in the description section of the video page. Users with internally facing videos can adapt this framework accordingly. Unlike an HTML transcript, a PDF can be posted just about anywhere.

Those who adopt and use the transcription system we have developed might consider putting a link to this paper in the Transcription Information section of each transcript. In public-facing videos, this might help expand awareness of our system and drive interest in further development. Additionally, in both internally facing and public-facing videos, a link to a publicly available paper signals that a sincere effort was made to produce a high-quality transcript with state-of-the-art mathematical accessibility. Naturally, we also hope to receive a measure of professional recognition in the event that our system is used or further developed by others.

### 5.11 Overall evaluation of v6m transcriptions

We think that the quality delivered by the v6m transcription workflow – if it includes the read-through and cleaning step described above – likely exceeds that which could be delivered by an expert human transcriber, unless they had a degree in physics and expert command of LaTeX or HTML and MathML. And while a human transcription expert with these further capabilities could likely do better with regard to descriptions of interactions with graphs, they would likely struggle to match our system's performance with regard to board-written equations.

However, most people who make instructional videos are unable to call upon the services of a professional transcriber, with or without disciplinary knowledge or accessible math coding expertise. They may not have access to any help at all, other than guides for HTML and MathML formatting or instructions for typing math into Canvas. Compared to the transcriptions or alternative video board notes that are likely to be produced manually under these real-world conditions, our system offers a vast performance upgrade and a much smaller time cost.

## 6 Neo-Minimalism: Exploring the Promise and Peril of Minimalist Instruction Sets

About two-thirds of the way into this project, we rechecked one of our very first test transcriptions, which had been carried out with a minimalist set of system instructions, for PDF/UA-2 and ISO 32005 compliance with PDFix, and to our surprise, it passed. We had originally deemed that transcription to be a failure; nobody had been able to give us reliable information about how to check a PDF, and we had not yet learned about PDFix. We had applied an incorrect test, and consequently we drew an incorrect conclusion at a very early fork in the evolutionary road.

Our belated realization that this early test had been a success rather than a failure led us to wonder about the path not taken. Could this path have led to higher performance with less evolutionary baggage? Could it still lead there? We decided to return to that early developmental fork and explore the minimalist branch; hence "neo-minimalism." Here, we summarize our development and testing of four iterations of neo-minimalist system instructions using video 3, and further testing of the most promising iteration (nm-2) using videos 1, 2, 7, and an older version of video 10. All four sets of neo-minimalist instructions and all resulting test PDFs are available in folder `c-neo-minimalism` in the supplementary materials [15].

In our first attempt (nm-1), we revised our earlier minimalist prompt, with AI assistance, to incorporate minimalist-compatible aspects of things we had learned. We continued to forbid the use of crash-provoking square brackets for visual descriptions and provided the most minimalist transition-signal alternative that we could imagine; these parameters were combined into the following audaciously minimalist instruction:

> Format speaker transitions cleanly as normal text paragraphs starting with "Instructor:". Do not use raw square brackets [ ] or parentheses ( ) for non-verbal cues, visual actions, or structural blocks.
>
> —`nm-1-sys-instructions.txt`, in [15]

The backward appearance of the double quotation mark before the word (Instructor) above is due to the use of typewriter quotes rather than unicode curly quotes in our system instructions in all but the latest stages of development. In the 6-series iterations, when we began instructing Gemini to use unicode curly quotes and apostrophes in its transcription output, we also began using unicode curly marks in our instructions to Gemini in order to set a good example.

To counterbalance the restriction imposed by the square-bracket prohibition and to unleash Gemini's creative capacity, we included the following "Operational Freedom" instruction:

> Outside of the structural and subscript constraints above, you have complete freedom to format, chunk, and typeset the spoken text and equations natively, using your best technical judgment for a clean, accessible physics transcript.
>
> —`nm-1-sys-instructions.txt`, in [15]

We also incorporated a streamlined version of our then-current subscript system, which still contained our earlier problematic "small math spaces" scheme for semantically separating subscript elements, to create version nm-1 of our neo-minimalist system instructions. We ran a test transcription on video 3 and pasted Gemini's output into the v6m preamble shell, which at the time we were calling v5a. The resulting PDF compiled without errors and passed the PDFix PDFUA-2-ISO32005 validation test, and Gemini's integration of board-written mathematics, carried out without any guidance other than the "Operational Freedom" instruction, was quite impressive. However, Gemini began every single presenter-spoken paragraph with an "Instructor" label. Additionally, for no reason that we can discern, Gemini added a hilariously inappropriate – if somewhat flattering – three-paragraph "Pedagogical Synthesis and Judgment" section; we quote the first paragraph below:

> Instructor Craig Looney's fully vectorized approach is pedagogically superior to the traditional method of components, as it shifts the cognitive load from tedious geometric projections to systematic vector algebra. This reduces student errors in sign conventions, which is a major bottleneck in introductory electromagnetism. Therefore, it is highly recommended to adopt this vectorized curriculum across all calculus-based physics sections starting this semester.

—`nm1-video-03.pdf`, in [15]

This experiment illustrates both the potential and the danger of minimally instructed Gemini. Encouraged by the potential, we subtly restructured the bracket-prohibition and the transition-signal instructions. In retrospect, we realized that a specific "instructor" label instruction would carry excessive force in an otherwise minimalist set of instructions, so we removed it. Instead, we clearly laid out our signaling intent, and gave Gemini even more latitude to carry it out. Here is the revised instruction in the nm-2 system instructions:

> Clearly and cleanly signal transitions between verbal transcription and non-verbal description in a way that is non-burdensome to screen-reader users and without using square brackets [ ] or parentheses ( ) which could cause fatal compilation errors.

—`nm-2-sys-instructions.txt`, in [15]

Rather than attempt to counter Gemini's post-transcript editorializing with a specific prohibition, we instead added a single sentence to the opening "Role and Target Environment" to more clearly lay out Gemini's purpose:

> Your **ULTIMATE OBJECTIVE** is to ensure that students reading the resulting PDF documents, including blind students using screen-readers, can perfectly reconstruct the spoken equations and the physical, visual arguments made by the presenter.

—`nm-2-sys-instructions.txt`, in [15]

These calibrated nudges produced a much-improved transcript with no post-transcript commentary that once again passed the PDFix validation test. Guided only by the general instruction to "signal transitions … " noted above, Gemini placed a "Visual description:" label at the beginning of every transition from verbal to nonverbal transcription. However, Gemini did *not* signal the subsequent return to verbal transcription with a corresponding "presenter" label (or any other label such as "speaker" or "instructor", etc.), which makes it substantially more difficult, even for a sighted reader, to recognize the transition back to verbal transcription.

In our next neo-minimalist iteration, nm-3, we tweaked the transition instruction to clarify that the return to verbal transcription also needed to be signaled; the `\verb|...|` command has been used to make the curly braces visible in the instruction below.

> Clearly and cleanly signal transitions from verbal transcription to non-verbal description – and from non-verbal transcription back to verbal transcription – with appropriate labels in a minimally burdensome way, so that blind students using screen readers can easily identify these transitions with minimal disruption. When signaling transitions, do NOT use square brackets [ ] or parentheses ( ) or curly braces `{ }` since all of these can cause fatal compilation errors (and since none of these provide blind readers with a clear transition signal).

—`nm-3-sys-instructions.txt`, in [15]

In response to this nudge, Gemini inserted ridiculously excessive transition labels; here is a representative example:

> — TRANSITION: VISUAL DESCRIPTION — The presenter points to the observation point $(0, -b, 0)$ on the $y$-axis. — TRANSITION: VERBAL TRANSCRIPTION —

—`nm3-video-03.pdf`, in [15]

There were further problems with the nm-3 transcription. Bold font was used, instead of arrows, to denote ordinary vectors. A single section heading was inserted at the beginning of the document, and was followed immediately by a single subsection heading; this is worse than having no section headings at all. While the resulting PDF passed the PDFix validation test, the overall result was a substantial step backward.

In what turned out to be our final neo-minimalist iteration (nm-4), we tried giving Gemini more structured guidance regarding presenter labels. In the following excerpt, LaTeX commands and curly braces have been made safely visible using `\verb|...|`; the stray quotation mark following the word (Presenter) in the 2nd sentence was a typo present in the actual instructions given to Gemini.

> Accessible segment labels: use `\bf{Visual description:}` to note the beginning of each non-verbal description segment, and use `\bf{Presenter:}` to note the RETURN to verbal transcription. Avoid disruptive repetition of the `\bf{Presenter"}` label within an ongoing verbal segment, even if the ongoing verbal segment contains LaTex math.) When signaling transitions, do NOT use square brackets [ ] or parentheses ( ) or curly braces `{ }` since all of these can cause fatal compilation errors (and since none of these provide blind readers with a clear transition signal).
>
> —`nm-4-sys-instructions.txt`, in [15]

While the entire transcript came out in bold font due to our mistaken use of `\bf{}` rather than `\textbf{}`, the revised instructions had the desired result relating to the placement and frequency of segment labels. Unfortunately, other problems emerged. Square brackets were used to inject short descriptions into verbal segments in spite of the clearly stated prohibition in the instructions. In the absence of a clear rule to avoid constructions that would be problematic for screen readers, Gemini used underbraces to literally mimic rather than accessibly transcribe board-written constructions. Finally, no section headings were inserted. In spite of these problems, the resulting PDF passed the PDFix validation test.

Although a minimalist approach may be possible, we saw no clear way forward, within a minimalist framework, to reliably prevent the nm-3 and nm-4 test issues from occurring in further iterations. While the earlier nm-2 test revealed tantalizing potential, we had serious doubts that we could ever achieve – using a minimalist instruction set – the reliable transcript quality that we had achieved with our main instruction set at that time, let alone the substantially higher performance we later attained with the v6m instructions that we ultimately developed.

Instead of attempting further neo-minimalist iterations, we tested nm-2 – the most promising neo-minimalist iteration – on videos 1, 2, 7, and an older version of video 10. The presentation quality of the resulting transcripts is quite good overall, except for the lack of a clear transition signal upon the return to verbal transcription. The transcriptions of videos 1 and 2 compiled without errors and passed the PDFix validation test. Video 7 had a compilation error flag but passed the PDFix validation test anyway. The transcription of the older version of video 10 compiled with errors and did not pass the PDFix validation test.

We believe that our neo-minimalist investigations suggest that it might be possible to develop a minimalist instruction set that reliably delivers high-quality transcription and error-free compilation into accessible PDFs, but that this is beyond our own prompt-engineering capabilities, at least at this time. Nevertheless, we hope that the experiments we have described and the instruction sets and test PDFs we have provided in the supplementary resources [15] will be useful to those who want to explore a minimalist approach.

## 7 Guide to Supplementary Materials

The durably archived supplementary materials [15] contain the following resources:

- `a-read-me.txt`: A repository information file describing the supplementary materials.
- `preamble-shell-v6m.tex`: A ready-to-use `.tex` shell document containing the v6m preamble.
- `system-instructions-v6m.txt`: The v6m system instructions and the 3-optioned kickoff prompt, for use with Gemini in the AI Studio environment.
- `z-gemini-thoughts-while-transcribing-video-11-v6m.txt`: Contains Gemini's thoughts while transcribing video 11 using the v6m system instructions (2026-07-16), copied and pasted into this file as text.
- (folder) `a-latex-files-v6m`: Contains the 17 LaTeX source files resulting from pasting in Gemini's v6m transcriptions of the 16 test videos into a shell document containing the v6m preamble (a.k.a. "version 6x").
- (folder) `b-pdf-files-v6m`: Contains the 17 PDFs produced by compiling the corresponding LaTeX source files with the LuaLaTeX engine.
- (folder) `c-neo-minimalism`: This folder contains files, in the subfolders noted below, relating to our relatively brief exploration and testing of neo-minimalist system instructions.
  - `neo-minimalist-system-instructions` (subfolder): Contains the four iterations of neo-minimalist system instructions (nm-1, nm-2, nm-3, nm-4) in `txt` file format.
  - `nm-1-2-3-4-video-03-test-pdfs` (subfolder): Contains the PDF files produced from single-shot transcriptions of video 3 using the four different sets of neo-minimalist system instructions.
  - `nm-2-tests-on-5-videos-pdfs` (subfolder): Contains the PDF files produced from single-shot use of the nm-2 system instructions on videos 1, 2, 3, 7, and an older version of video 10.

We hope these materials will be useful for those who want to use, study, or further develop our transcription system.

## 8 Conclusion

We have developed, tested, and presented an AI-powered workflow that uses freely available tools to transcribe introductory physics videos into PDF transcripts featuring state-of-the-art mathematical accessibility. The test PDFs and LaTeX source files described above – and durably archived in the publicly available supplementary materials [15] – demonstrate the extraordinary potential of AI transcription. We hope that others will build upon our work and also develop alternative AI transcription frameworks to leverage the ever-growing capabilities of multimodal generative AI with ever-greater effectiveness. In principle, it should be possible to translate our workflow into a one-button system that could spit out, at the user's option, a compiled PDF or an editable LaTeX source file. In the meantime, our v6m system instructions and paired preamble allow instructors, right now, to rapidly produce high-quality PDF transcripts with gold-standard mathematical accessibility, provided they are willing and able to work with Gemini in the AI Studio environment and deal with a local-machine LaTeX installation. We suggest that modest institutional investments in centralized LuaLaTeX expertise have the potential to vastly reduce the overall cost and human effort of the enormous lift that will be necessary to achieve widespread accessible transcription of videos with substantial mathematical content. Finally, we hope that this paper will lead to wider appreciation of the fact that LuaLaTeX-produced SE-tagged PDFs offer the same level of technical mathematical accessibility as HTML plus MathML in a substantially more stable and portable format.

## Declaration of Use of Generative AI and AI-Assisted Technologies

The authors utilized generative AI technologies in the following research, development, and manuscript preparation aspects of this project:

- **Test transcripts:** Consistent with the purpose of this project and as described in the manuscript, Gemini 3.5 Flash was used in producing all test transcripts discussed in the manuscript and archived in the supplementary materials [15].
- **Development of preamble and system instructions:** As described in Section 3, Gemini models were extensively used in developing and refining the LaTeX shell-document preamble code and the transcription system instructions.
- **Drafting the Introduction:** Gemini 3.5 Flash was provided with advanced manually written drafts of the abstract and manuscript body, and prompted with author-provided parameters to write an initial draft of the Introduction. While most of this draft was discarded, the structure and wording of the first paragraph of the finished Introduction draw heavily on the AI-generated draft. Gemini was also used heavily for wording, phrasing, and refinement suggestions – always subjected to critical author evaluation – during the drafting phase of the author-framed second paragraph. The first sentence of the third paragraph of the Introduction was adapted from Gemini-generated wording. The rest of the Introduction was drafted manually.
- **Language editing and refinement:** Beyond the drafting of the Introduction, AI was used throughout the manuscript for standard spelling and grammar checks and for wording and phrasing refinement consultations. All AI-generated suggestions were critically evaluated by the authors, and all actual manuscript edits were carried out manually. No AI model was ever granted direct editing access to the manuscript file.

The authors have carefully reviewed all aspects of the manuscript and take full responsibility for its contents and the integrity of the published work.